А.С. Смирнов, К.С. Голохваст, А.В. Тумялис

СМИРНОВ АЛЕКСЕЙ СЕРГЕЕВИЧ – младший научный сотрудник, smirnov.aserg@dvfu.ru
ГОЛОХВАСТ КИРИЛЛ СЕРГЕЕВИЧ – директор, golokhvast.ks@dvfu.ru
ТУМЯЛИС АЛЕКСЕЙ ВЯЧЕСЛАВОВИЧ – старший научный сотрудник, tumialis.av@dvfu.ru

Дальневосточный региональный научный центр Российской академии образования

Дальневосточный федеральный университет

690922, Россия, Приморский край, о. Русский, кампус ДВФУ

# Развитие «Интернета вещей», дополненной реальности и коммуникационных технологий (обзор)

Аннотация: Подобно тому, как появление персональных компьютеров и смартфонов изменило жизнь современного общества, «Интернет вещей», дополненная реальность и сверхбыстрые и надежные телекоммуникационные сети нового поколения, путем объединения физических объектов реального мира с постоянно растущей вычислительной мощью и интеллектом киберпространства, способны совершить следующую большую революцию во всех сферах деятельности человека будущего.

Ключевые слова: Интернет вещей, 5G, дополненная реальность.

# ВВЕДЕНИЕ

Интернет вещей (IoT) - это новая и многообещающая технология, нацеленная на глобальное преображение нашего мира путем объединения физических объектов («вещей») и киберпространства. Концепция IoT объединяет в себе множество технологий, таких как Интернет, распределенные вычисления, машинное обучение, коммуникации, большие данные, сенсорные технологии, человеко-компьютерное и межмашинное взаимодействие.

Интернет вещей, аналитика больших данных и машинное обучение - развивающиеся области науки и техники, формирующие новое поколение компьютерных систем на основе использования искусственного интеллекта. Важно отметить, что эти области междисциплинарны по своей природе, что позволяет им расти как с точки зрения их теоретических основ, так и с точки зрения их практического применения [1].

# «ВЕЩИ» В «ТУМАНЕ»

Кирпичиками, из которых строится «Интернет вещей», являются «умные» вещи. Наделенные интеллектом повседневные объекты превращаются в «умные» вещи, способные не только собирать информацию из окружающей среды и взаимодействовать с физическим миром, но и связываться друг с другом через коммуникационные каналы связи для обмена данными и информацией. Ожидаемое огромное количество взаимосвязанных устройств и значительный объем доступных данных открывают новые возможности для создания услуг, которые принесут ощутимую пользу обществу, окружающей среде, экономике и отдельным людям [2]. Конечной целью воплощения концепции IoT является обеспечение того, чтобы носимые устройства, датчики, стиральные машины, планшеты, смартфоны, интеллектуальные транспортные системы и многие другие объекты были подключены через Интернет к общему интерфейсу с возможностью общения друг с другом и могли взаимодействовать без вмешательства в их работу человека. Воплощение концепции «Интернета вещей» даст любому человеку со смартфоном или другим мобильным устройством контролировать что угодно в своем окружении [3].

В зависимости от того, в какой сфере применяют «Интернет вещей», различают Массовый IoT и критический IoT [4]. Первый задействован для работы потенциально огромного количества, порядка десятков миллиардов устройств, тогда как второй реализуется там, где важна максимальная скорость и доступность в паре с минимальной задержкой. Ожидается, что IoT создаст благоприятную среду, которая будет влиять на нашу повседневную жизнь и на бизнес-среду и способствовать росту мировой экономики через Массовый IoT и критический IoT, в зависимости от характера применения этой технологии. Массовый IoT требует подключения огромного числа интеллектуальных устройств, которые могут быть развернуты в логистике, «умных домах» и «умных городах», интеллектуальных энергосистемах, аграрном мониторинге и т.д. Например, для некоторых людей уже стало обыденностью когда «умный дом» сам открывает ворота гаража перед машиной хозяина, управляет освещением и климатом в комнатах и заказывает доставку продуктов к ужину. Устройства подобного «умного дома» должны быть недорогие, с низким энергопотреблением, расширенной зоной покрытия и высокой масштабируемостью для эффективного развертывания Массового IoT. С другой стороны, устройства Критического IoT, например, удаленно управляющие аппаратом искусственной вентиляции легких пациента и следящие за его жизненными показателями требуют более высокой скорости передачи данных, надежности, безопасности и как можно меньшей задержки, поскольку отказ таких устройств может привести к серьезным последствиям [5].

Предполагается, что количество подключенных к «Интернету вещей» устройств к 2025 году превысит 7 триллионов, при расчете примерно 1000 устройств на человека. Часть из них будет носимыми гаджетами, но большинство будет задействовано в инфраструктуре [2]. Это простые датчики, измеряющие скорость, температуру, давление и другие физические параметры, которые только снимают данные и передают информацию для обработки в «облако» – удаленные порой на тысячи километров от места действия датчика дата-центры. По мере роста количества таких объектов и совершенствования технологий объем производимых ими данных будет увеличиваться. В дополнение к увеличению объема, эти большие данные также характеризуются локализацией в определенном пространстве, временными параметрами, а также разным качеством. Интеллектуальная обработка и анализ этих больших данных являются ключом к разработке умных приложений IoT [6].

Внедрение IoT требует новой инфраструктуры обработки огромного объема генерируемых миллиардами устройств данных и превращения их в знания для дальнейшего использования. Современные облачные модели не предназначены для такого объема, разнообразия и скорости данных, которые создает IoT. Миллиарды ранее не подключенных устройств генерируют более двух эксабайт данных каждый день. Примерно 50 миллиардов «вещей» будут подключены к Интернету к 2020 году. Перемещение всех данных из этих «вещей» в «облако» для анализа потребует огромных объемов пропускной способности [7].

Устройства IoT генерируют данные постоянно, и часто анализ этих данных должен быть очень быстрым. Например, когда температура в химической ванне быстро приближается к приемлемому пределу, корректирующие действия должны быть предприняты почти сразу же. В то время, когда показания температуры перемещаются от датчика к «облаку» для анализа, возможность предотвратить нештатную ситуацию и спасти партию продукции может быть уже потеряна. Для обработки такого объема, разнообразия и скорости данных IoT требуется новая вычислительная модель.

Таким образом, простая двухуровневая архитектура «Облако – Объект» вряд ли может обеспечить качественную коммуникацию и обработку данных, необходимые для всех этих миллиардов подключенных устройств. Подключение множества датчиков непосредственно к «облаку» чрезвычайно требовательно к облачным ресурсам. В результате «облако» остается занятым в течение каждого рабочего цикла датчика. Кроме того, пропускная способность не может поддерживать эту загрузку данных [8]. В дополнение к облачной инфраструктуре обработки и хранения данных, предлагается архитектура туманных вычислений. Туманные вычисления были разработаны, чтобы располагаться между объектами IoT и «облаком», обеспечивая иерархию вычислительной мощности, которая может собирать, объединять и обрабатывать данные с устройств IoT. Распределение этой иерархии может варьироваться от места к месту, но предполагается, что первый уровень будет расположен на расстоянии одного шага от периферийного устройства, в точке доступа (например, WiFi-роутер или сотовый телефон) или сразу над ней. Это будет первый (ближайший) вариант разгрузки для периферийных устройств, обеспечивающий меньшую задержку даже при ограниченной вычислительной мощности [9], ведь сами объекты IoT обладают вычислительной мощностью и оперативной памятью. Эти устройства, промышленные контроллеры, коммутаторы, маршрутизаторы, встроенные серверы и камеры видеонаблюдения, называемые узлами «тумана», могут находиться в любом месте с сетевым подключением: в заводском цехе, в верхней части опоры линии электропередач, вдоль железнодорожного пути, в транспортном средстве или на нефтяной вышке. Любое устройство с подключением к компьютерам, хранилищам и сетям может быть узлом «тумана».

Проводя расчеты в «тумане», как можно ближе к периферийным устройствам IoT, объем данных, который требуется переместить в «облако» для обработки, значительно снижается. Также обеспечивается минимальная задержка обработки данных, реализуется принцип локализации данных на местности [10] и экономится заряд батареи смартфона или AR-очков. Вычисления, проводимые в архитектуре «тумана», также предлагается делить на классы [11]. Эти классы могут использоваться для определения приоритетов сетевого трафика и требований к обработке планировщиками и механизмами распределения ресурсов. Приоритизация будет поддерживать обработку чувствительных к задержке приложений. Содействуя распределению нагрузки между слоями «тумана», наиболее вероятно, что туманные вычисления смогут поддерживать большее количество запросов, способствуя их масштабируемости. Комбинация «тумана» и «облака» может уменьшить объем передаваемых данных и убрать эффект бутылочного горлышка при их передаче в «облако», а также способствовать снижению задержек, так как ресурсы «тумана» задействуются непосредственно ближе к периферии IoT. В симуляционной модели было проанализировано [12] использование туманных вычислений совместно с облачными. Решение было приблизительно на 2.72% более энергоэффективно в сравнении с использованием только облачных вычислений.

Размещенная на сайте компании Cisco Systems, Inc. статья [13] описывает ряд требований к архитектуре туманных вычислений:

- Минимизация задержки: миллисекунды имеют значение, когда вы пытаетесь предотвратить остановку производственной линии или восстановить электроснабжение;
- Сохранение пропускной способности сети: нецелесообразно переносить огромные объемы данных с тысяч или сотен тысяч периферийных устройств в «облако». В этом нет необходимости, поскольку многие критические анализы не требуют облачной обработки и хранения;
- Решение проблем безопасности: данные IoT должны быть защищены как во время передачи, так и во время хранения. Это требует мониторинга и автоматического реагирования по всему континууму возможной кибер-атаки: до, во время и после;
- Надежная работа: данные IoT все чаще используются для принятия решений, влияющих на безопасность граждан и критически важную инфраструктуру. Целостность и доступность данных не может быть под вопросом;
- Сбор и защита данных в различных условиях окружающей среды: устройства IoT могут быть распределены на территории сотни и более квадратных километров. Устройства, развернутые в суровых условиях, таких как автомобильные дороги, железные дороги, полевые подстанции и транспортные средства, могут нуждаться в повышенной прочности и стойкости к атмосферным явлениям;
- Перемещение данных в лучшее место для обработки: какое место лучше, зависит от того, насколько быстро требуется решение. Чрезвычайно чувствительные ко времени решения должны быть приняты как можно ближе к «вещам», производящим и действующим на полученных данных. Напротив, для анализа больших данных, не требовательных к скорости обработки, годятся вычислительные мощности «облака».

Обработка данных в «тумане», ближе к месту их производства и востребованности, решает проблемы, связанные с увеличением объема, разнообразия и скорости обработки данных. Туманные вычисления ускоряют обнаружение и реакцию на события, исключая ненужное движение данных в «облако» и обратно. Это устраняет необходимость в дорогостоящей модификации сети для добавления пропускной способности. Туманные вычисления также защищают конфиденциальные данные IoT, обрабатывая их внутри стен компании. В конечном итоге организации, использующие туманные вычисления, получают повышение гибкости бизнеса, уровня обслуживания и безопасности. Для подключения устройств IoT к «туману» обычно используются беспроводные каналы связи, поскольку эти устройства часто имеют только беспроводные интерфейсы. Беспроводные соединения могут также использоваться в соединениях между «туманом» и

«туманом» в зависимости от доступной инфраструктуры. Ожидается, что технологии мобильной связи (3G, 4G, 5G) будут задействованы в системах туманных вычислений.

Будущее IoT связывают с развитием телекоммуникационных сетей пятого поколения [14], поскольку 5G позволит развивать скорость загрузки 20 Гбит/сек в сети мобильной связи при 4 миллисекундах задержки (в зависимости от среды распространения и используемой частоты сигнал приходит к получателю с той или иной задержкой; поколение связи 4G работало при 10-50 мс задержки). Текущие разработки в области 5G характеризуются организацией сетевых ресурсов для удовлетворения широкого спектра услуг, классифицированных по трем категориям. Расширенная широкополосная мобильная связь (enhanced mobile broadband - eMBB) должна будет обеспечивать беспроводную передачу огромных объемов данных на устройства виртуальной и дополненной реальности, в телемедицине и при трансляции видео сверхвысокого разрешения. Сверхнадежная связь с малой задержкой (ultra-reliable and low-latency communications - URLL) будет задействована в автопилотируемых машинах и летательных аппаратах. Массивная межмашинная связь (massive machine-type communications - mMTC) обеспечит функционирование умных городов и домов, умной инфраструктуры и умных автоматических систем промышленного мониторинга [15].

Хотя иерархия туманных и облачных вычислений применяется во всех трех данных категориях, услуги IoT обычно рассматриваются в контексте mMTC. Примечательная особенность организации сетей пятого поколения - это так называемый слайсинг сети. Оператор, предоставляющий услуги 5G, сможет разделить всю сеть на доли (слайсы), предоставляя каждую долю для удовлетворения самых разных потребностей вышеобозначенных трех категорий услуг в зависимости от требований к задержке, надежности, пропускной способности, масштабируемости и поддержки мобильности. В этом контексте долю сети можно рассматривать как виртуальную сеть, ресурсы которой предоставляются для конкретной услуги или класса услуг и изолируются от других долей, которые совместно используют одну и ту же физическую инфраструктуру.

В результате получается более гибкая, надежная, масштабируемая и безопасная сеть. Используя эти технологии, во многих ситуациях сети смогут переконфигурировать доли в течение нескольких секунд, чтобы быстро реагировать на местные требования, такие как неожиданное собрание людей или приоритизация систем аварийных служб. С другой стороны, также можно запрограммировать долгосрочную аренду доли, например, для компании, занимающейся электроснабжением, для размещения ее компонентов, таких как счетчики, датчики, контроллеры и другие устройства IoT. Краткосрочная аренда также возможна, например, когда организаторы концерта хотят иметь выделенную долю для музыкального фестиваля на выходных и оптимизировать его для потоковой передачи высококачественных видео и музыкальных данных [9]. Ожидается, что внедрение сетей пятого поколения (5G) еще больше ускорит повсеместное внедрение «Интернета вещей» и повысит его удобство использования и безопасность [16].

Лавинообразно растущие данные, даже после обработки в «тумане» и «облаке» и превращения в знания, потребуют новые инструменты для работы с ними. Смартфоны, очки дополненной реальности и другие носимые гаджеты генерируют огромное количество данных для IoT, поэтому эти данные должны быть доступны в режиме реального времени. Кроме того, устройства должны воспринимать информацию из окружающей среды и передавать ее непосредственно в облачную аналитическую систему. Принимая это во внимание, можно сделать вывод, что облачные хранилища и обработка данных в «тумане» и «облаке» значительно расширяют возможности умных носимых устройств.

## ИНДУСТРИЯ 4.0

Объединение уже существующих промышленных автоматических систем с большими данными, вычислительной мощью, скоростью и надежностью «Интернета

вещей» может позволить достичь четвертой промышленной революции, часто называемой просто «Индустрия 4.0» [17]. В обозримом будущем на умных заводах будут использоваться мониторинг состояния всего производственного оборудования в режиме реального времени, управление режимом работы на основе облачных вычислений и сервис-ориентированных технологий, проектирование, тестирование, сборка и обслуживание продукции используя методы виртуальной/дополненной реальности [15].

Помимо того, высококонкурентная бизнес-среда требует инновационных продуктов при минимизации времени их выхода на рынок, и также растет тренд на создание совместных производственных сред, в которых различные заинтересованные стороны на всем протяжении жизненного цикла разработки продукта в реальном времени обмениваются друг с другом информацией. Дополненная реальность, представляющая собой набор инновационных и эффективных методов взаимодействия человека и компьютерными сетями, обладает потенциалом для решения этих проблем [18]. Дополненная реальность это одна из ключевых стимулирующих технологий, которая позволит передавать информацию о производстве и обслуживании от цифровых систем проектирования и баз данных операторам на месте. Дополненная реальность в паре с передовыми коммуникационными технологиями, поддерживающими высокоскоростную передачу данных и сочетающих в себе низкую задержку и повышенную доступность, такими как сети 5G, предлагают способ оперативного распространения информации и повышения осведомленности о состоянии в «умном» фабричном цехе [19], за счет чего повысится безопасность производства [20]. Предприятия, внедрившие технологии «Интернета вещей» и дополненной реальности, смогут значительно увеличить свою производительность, конкурентоспособность и прибыльность за счет сокращения жизненного цикла разработки и производства продукта, а также за счет наличия в режиме реального времени точной информации из «Интернета вещей» о том, что, как и в каких количествах нужно произвести [21]. Используя помощь устройств дополненной реальности, операторы «умных» фабрик смогут производить высококастомизированную продукцию, удовлетворяющую самым изысканным запросам потребителя, массовое производство останется в прошлом [22].

Сельское хозяйство также будет умным и точным, так как IoT вносит значительный вклад и в эту область, давая фермерам возможность контролировать свои фермы дистанционно и более эффективно руководить сельскохозяйственной деятельностью, помогая в реализации концепции точного земледелия. Точное земледелие [23] позволяет фермеру получать увеличенный урожай за счет лучшего управления, а применение более подходящих или щадящих химических удобрений и пестицидов только там, где это действительно необходимо, также способствует сохранению окружающей среды [24].

## УМНЫЕ ГОРОДА БУДУЩЕГО

Но в первую очередь, развитие и повсеместное распространение технологий IoT, VR/AR и 5G коснется каждого из нас, до неузнаваемости изменив привычную нам повседневную бытовую среду. Происходившая в последние два десятилетия технологическая революция, обусловленная достижениями и разработками в области информационных и коммуникационных технологий, кардинально изменила то, как мы общаемся, работаем, путешествуем, живем. Наши города должны стать интеллектуальной динамичной инфраструктурой, которая будет удовлетворять все запросы граждан, при этом выполняя критерии энергоэффективности и надежности.

«Умный город» - это городская система, которая использует такие информационные и коммуникационные технологии, как IoT, VR/AR и 5G, чтобы сделать свою инфраструктуру и общественные услуги более интерактивными, более доступными и более эффективными для своих жителей.

Выделяют [25] следующие компоненты «умного города»:

- «Умная» инфраструктура сетей водного и энергоснабжения с датчиками контроля расхода ресурсов;
- «Умные» транспортные сети со встроенными системами мониторинга и контроля в реальном времени;
- «Умная» среда, реализующая принцип защиты и надзора природных ресурсов, например, система управления отходами и контроль загрязнения окружающей среды с помощью датчиков;
- «Умные» услуги здравоохранения, образования, туризма, безопасности и т.д.;
- «Умное» управление городом, связанное с технологиями предоставления умных услуг и использованием ресурсов в соответствии с выбранной политикой;
- «Умная» экономика, способствующая росту бизнеса и занятости населения, а также расширению города;
- «Умный» образ жизни, который улучшает качество жизни в городской среде;
- И, конечно же, умные люди, которые творят и внедряют инновации в этом высокотехнологичном обществе.

Появление «умных» городов обусловлено, главным образом, тремя вызовами: увеличением населения Земли и ростом миграции населения из сельских регионов в городские центры (прогнозируется, что к 2050 году доля городского населения достигнет 70%), нехваткой природных ресурсов, а также загрязнением окружающей среды и изменением климата. Обществу необходимо решать эти задачи, чтобы минимизировать потребление невозобновляемых энергетических ресурсов, продвигать использование возобновляемых источников энергии и сокращать выбросы $CO_2$ в атмосферу в этих густонаселенных городах и новых городских районах. Концепция «Умный город» является мощным инструментом для решения этих вызовов, удовлетворяя при этом потребности города и его жителей [26]. Для полноценного использования всех сервисов и объектов IoT, которые есть в «умном городе», будут использоваться смартфоны, очки дополненной реальности и другие «умные» носимые устройства. Через них будет собираться информация об активности жителей, чтобы сделать важнейшие компоненты инфраструктуры и услуг города более интерактивными, доступными и эффективными. По полностью безопасным и лишенным привычных заторов дорогам будут перемещаться автономные автомобили [27], но если пассажир решит взять управление в свои руки, то проецирование навигационных данных через устройство дополненной реальности на лобовом стекле, взаимодействующее с установленными в машине и на местности датчиками, поможет ему найти путь в незнакомом районе [28]. Применение дополненной реальности в медицинской сфере, особенно в хирургии для тренировок стажеров и при проведении различных операций [29, 30, 31], уже показало свой положительный эффект на точность и скорость выполнения многих процедур, что в конечном итоге позитивно сказывается на состоянии пациента, и в больницах умного города тем более будет расти потребность в комбинации технологий AR и IoT для обеспечения конкурентного преимущества в существующих хирургических технологиях [32]. За пределами медицинского учреждения с помощью устройств AR через IoT станет возможен постоянный удаленный мониторинг людей с различными недугами со сверхточным и сверхбыстрым анализом данных и с напоминаниями для пациента, какие медикаменты ему необходимо принять, тем самым, корректировку лечебного плана в зависимости от изменившихся внешних условий можно будет делать в режиме онлайн.

Революционные изменения из-за применения технологий AR и IoT происходят в сфере торговли, где любая даже самая незначительная инновация дает неоспоримые конкурентные преимущества над соперниками в стремлении в полной мере удовлетворить все запросы покупателей. Инновации внедряются как со стороны спроса, так и предложения. Повсеместное использование недорогих, не требующих источника питания радио-меток приводит к повышению эффективности работы логистических каналов за счет анализа складских остатков в реальном времени. Многие производители одежды и обуви уже перешли на установку таких меток непосредственно сразу на фабрике [33]. Считывание метки AR-устройством потенциального покупателя не только представляет ему полнейшую информацию о товаре, но и позволяет реализовать новые способы продвижения товара средствами дополненной реальности. Расположенные в торговом зале датчики Интернета вещей являются предметом повышенного интереса для маркетологов, так как позволяют собирать массу данных о посетителях и получать знания о покупательских паттернах и предпочтениях, оптимизируя расположение товаров в магазине и создавая персонализированные предложения для каждого покупателя. Из магазинов одежды исчезают примерочные, ведь приглянувшуюся деталь гардероба можно примерить на свое отражение в интерактивном зеркале с помощью средств AR. Объединение этих инновационных технологий в высококонкурентной сфере торговли кардинально изменит шоппинг. Примером такого радикального изменения могут служить магазины Amazon, в которых нет ни касс, ни продавцов, ни консультантов, Покупатель идентифицируется в системе камерами наблюдения, берет понравившийся товар с полки и просто выходит из магазина, а оплата за покупку списывается с его счета.

## ЗАКЛЮЧЕНИЕ

Можно сделать вывод, что несмотря на то, что мир эволюционирует и становится все сложнее, появляются новые сложные высокотехнологичные системы, взаимодействующие с другими еще более сложными системами, но, тем не менее, вместе с этим, возникают абсолютно новые сферы человеческой деятельности, а существовавшие ранее радикально изменяются в лучшую сторону. В статье были поверхностно рассмотрены только некоторые частые примеры реализации функционирования IoT, дополненной реальности и телекоммуникационных технологий нового поколения. Каждая из этих технологий по отдельности имеет большой потенциал для того, чтобы сделать жизнь людей более безопасной, удобной и радостной, а при достижении их синергетического взаимодействия положительный эффект будет просто колоссальный. Области применения этих новых технологий абсолютно любые и ограничиваются только человеческой фантазией.